\documentclass[slac_one]{revtex4}
\usepackage{amsmath,amsfonts,amssymb}
\usepackage{fancyhdr}
\usepackage{ifpdf}
\ifpdf
  \usepackage{color}
  \definecolor{darkblue}{rgb}{0.3,0.3,0.6}
  \usepackage[
    pdftitle={Statistical analysis of a subset of the string theory landscape},
    pdfauthor={Florian Gmeiner},
    pdfsubject={String phenomenology},
    pdfkeywords={string landscape, vacuum statistics, type II string theory, ICHEP08},
    colorlinks=true,linkcolor=black,urlcolor=darkblue,filecolor=darkblue,
    citecolor=black,menucolor=darkblue,breaklinks=true,bookmarks=true,
    anchorcolor=black,unicode=true
  ]{hyperref}
  \usepackage[pdftex]{graphicx}
\else
  \usepackage{graphicx}
  \def\href#1#2{#2}
\fi

\def\aref#1#2{\ifpdf\href{http://arxiv.org/abs/#1}{arxiv:#1\if#2\else [#2]\fi}\else arxiv:#1\if#2\else [#2]\fi\fi}

\def\rotimg#1{\includegraphics[height=.25\textwidth,angle=90]{#1}}
\def\fourimages#1#2#3#4#5#6{\begin{figure}[th]\rotimg{#1}\rotimg{#2}\rotimg{#3}\rotimg{#4}\caption{#5}\label{#6}\end{figure}}

\begin{document}

\title{Statistical analysis of a subset of the string theory landscape}

\author{Florian Gmeiner}
\affiliation{NIKHEF, Kruislaan 409, 1098 SJ Amsterdam, The Netherlands}

\begin{abstract}
An analysis of a special class of type II string theory compactifications
is presented. We focus on recent work in one particular orientifold
background of intersecting brane models and the resulting
four dimensional gauge group and matter content. Special
attention is paid to solutions that resemble the gauge group of the MSSM.
Statistical correlations between properties of the models are analysed
and compared to results on different backgrounds.
\end{abstract}

\begin{flushright}\small NIKHEF/2008-028\\October 20, 2008\end{flushright}

\maketitle

\thispagestyle{fancy}

\section{INTRODUCTION}
Understanding the structure of string theory compactifications to four
dimensions and thereby finding a way to make contact with the Standard Model
of particle physics (SM) is the main task of string phenomenology.
There are two general ways to approach this problem. One approach is to
engineer the SM (or its supersymmetric variant, the MSSM) as closely as
possible in one particular setup. Alternatively one can try to analyse larger
classes of compactifications as a whole, using statistical methods to classify
properties of the low energy theories.
The latter approach has recieved quite some attention in the recent years,
in particular since it has become clear that the number of string theory vacua,
also known as ``the landscape'', is extremely large, even after including the
effects of moduli stabilization.
Interesting questions that can be asked in this context include the
search for common properties within the ensemble of models, in particular
the occurance of the gauge groups and matter content of the MSSM or
simple Grand Unified Theories (GUTs).
Another promising approach relies on the search for correlations
between properties of the models~\cite{stat1,corr}. Finding such correlations
in different corners of the landscape would be very exciting and might
finally even lead to predictions for the experimental search for physics
beyond the SM.

In the following we discuss some of the recent progress in understanding
the open-string part of the landscape. In particular we will focus on one
specific construction, intersecting brane models (IBMs) of type IIA string
theory on an orientifold background $T^6/\mathbb{Z}_6'$, where some
interesting structure has been found recently~\cite{z6p,gh07,gh08}.
IBMs are a very nice laboratory to study the gauge sector of string
compactifications, because their properties can be formulated in terms
of simple algebraic equations. This simplifies the task to formulate
computer algorithms in order to obtain the low energy spectrum of these models,
which makes it possible to analyse large ensembles of compactifications
on various orbifold backgrounds. In particular there has been work on
$T^6/(\mathbb{Z}_2\!\times\!\mathbb{Z}_2)$~\cite{stat1,z2z2,z2c,dt} and
$T^6/\mathbb{Z}_6$~\cite{z6}, which differs from $\mathbb{Z}_6'$ only by
the way the discrete group acts on the geometry.
These two cases will be used
to compare results on correlations with the $\mathbb{Z}_6'$ models.
Moreover we will use results obtained for Gepner models~\cite{gepner}
to compare them with the IBM case.

\section{SETUP}
For the sake of brevity we will give only a brief summary of the most
important features of IBMs. For more detailed information we refer the
reader to one of the reviews on the subject, a recent one is~\cite{ibmrev}.

We study compactifications of type IIA string theory on the background
$M=T^6/\mathbb{Z}_6'$. An orientifold projection, accompanied by an
involution in the compact space, introduces six dimensional
orientifold planes (O6-planes) carrying RR-seven-charge. D6-branes are
introduced to cancel this charge on $M$.
Since we are interested in supersymmetric compactifications, these branes
have to wrap calibrated (special Lagrangian) three-cycles in the orbifold.
The possible cycles can be paramterized by a choice of basis in the
homology $H_3(M,\mathbb{Z})$, hereby introducing some oversimplification,
since we should in fact use K-theory to describe the D-branes accurately.
Therefore we have to introduce an additional consistency condition, which
can be formulated in our setup by a sum over intersection numbers of
probe branes along the directions of the O6-planes.

A detailed description of the $T^6/\mathbb{Z}_6'$ setup can be found
in~\cite{gh07,gh08}, the main features are the following.
Due to the action of the orbifold group on the factorized $T^6$, described by complex coordinates $z^k$,
$\theta: z^k \rightarrow e^{2 \pi i v_k} z^k$ with
$\vec{v}=\frac{1}{6}(1,2,-3)$
and the geometric part of the orientifold projection,
$\mathcal{R}: z^k \rightarrow \overline{z}^k$,
the $SU(3)\times SU(3)\times SU(2)^2$ torus lattice can have one of two
possible orientations {\bf A} or {\bf B} on the $SU(3)$ parts and
{\bf a} or {\bf b} on the $SU(2)^2$ part.
The three-cycles on this orbifold can be split into a four-dimensional
part with basis $\rho_i$, $i=1,\ldots, 4$ inherited from the torus and
an eight-dimensional part with basis $\delta_j,\tilde{\delta}_j$, $j=1,\ldots, 4$ associated with the $\mathbb{Z}_2$ sub-symmetry.
The remaining 12 cycles corresponding to the $\mathbb{Z}_3$ sub-symmetry will not concern us here.
A general fractional cycle can be written with paramters $\tilde{a},d,e$ as
\begin{equation}\label{eq:brane}
 \Pi^{\rm frac} = \frac{1}{2} \Pi^{\rm bulk} +\frac{1}{2}\Pi^{\rm ex} = \frac{1}{2}\left( \sum_{i=1}^4 \tilde{a}_i \rho_i +\sum_{j=1}^4 \left( d_j \delta_j +e_j \tilde{\delta}_j \right)\right),
\end{equation}
where we introduced the notation bulk and ex(ceptional) cycles for the torus and $\mathbb{Z}_2$ cycles, respectively.

The RR tadpole cancellation condition and the supersymmetry
conditions for the D-branes can be written as
\begin{equation}\label{eq:cond}
\sum_a N_a \vec{X}_a = \vec{L}\quad\mbox{and}\quad
X_1+c_1X_2 > 0,\quad Y_1+c_2Y_2=0,
\end{equation}
where $N_a$ is the number of branes on stack $a$, $\vec{L}=(L_1,L_2,0,0,0,0)$ gives the orientifold charge and the coefficients $\vec{X}$ and $\vec{Y}$ are given explicitly in table~1 and appendix~D of~\cite{gh08}.
The coefficients $c_1$ and $c_2$ depend on the complex structure modulus $\rho=\sqrt{3}R_2/(2R_1)$ and the orientations as
$\vec{c}=(2\rho,3/(2\rho))$ for {\bf AA} and {\bf BB} geometries and $\vec{c}=(2\rho/3,1/(2\rho))$ for {\bf AB} and {\bf BA} geometries.
The exceptional part of the tadpole cancellation condition for fractional cycles
does not recieve any contribution from the orientifold planes.
The supersymmetry condition is fulfilled if~\eqref{eq:cond} holds for the bulk
part and the exceptional cycles wrap fixed points traversed by the bulk cycles.
The aformentioned K-theory condition does not constrain the solutions
further, which is a special feature of this orbifold and has been proved in~\cite{gh07}.

The full matter spectrum for these models can be casted in algebraic form.
This has been shown in~\cite{gh07} and confirmed using a different method in~\cite{gh08}.
In the case of branes intersecting at three non-vanishing angels it is given by
\begin{equation}\label{eq:spec}
\chi^{ab} = -\frac{1}{2}\sum_{k=0}^{N-1} \left(I_{a (\theta^k  b)} + I^{\mathbb{Z}_2}_{a (\theta^k b)}\right) \quad\mbox{and}\quad
\varphi^{ab} = \frac{1}{2}\sum_{k=0}^{N-1} \left| I_{a (\theta^k b)} + I^{\mathbb{Z}_2}_{a (\theta^k b)}  \right|,
\end{equation}
where $\chi$ gives the chiral and $\varphi$ the complete spectrum.
The intersection numbers between bulk and exceptional branes
and their orbifold images generated by $\theta$ can be computed
using the coefficients in the expansion~\eqref{eq:brane}.
In a similar fashion the spectrum of symmetric, antisymmetric and adjoint representations of the gauge group can be computed.
In the cases where one or more of the angles between the branes are vanishing
or the branes lie on top of the orientifold planes, the formula has to be slightly modified. Details can be found in~\cite{gh07,gh08}.

\section{STATISTICAL DISTRIBUTIONS}
\subsection{General statistics}

\begin{figure}[th]
\includegraphics[width=.33\textwidth]{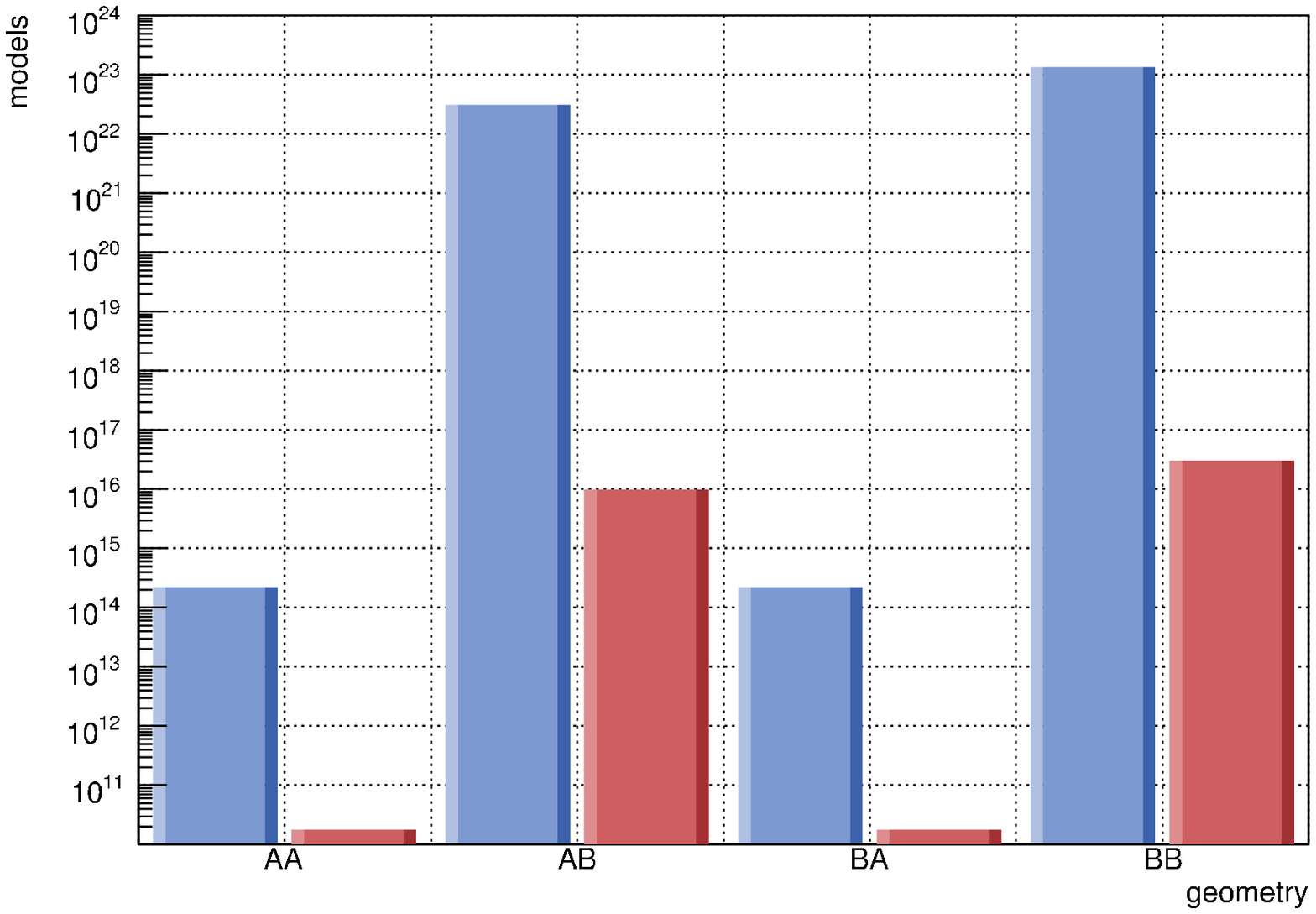}\hfill
\includegraphics[width=.33\textwidth]{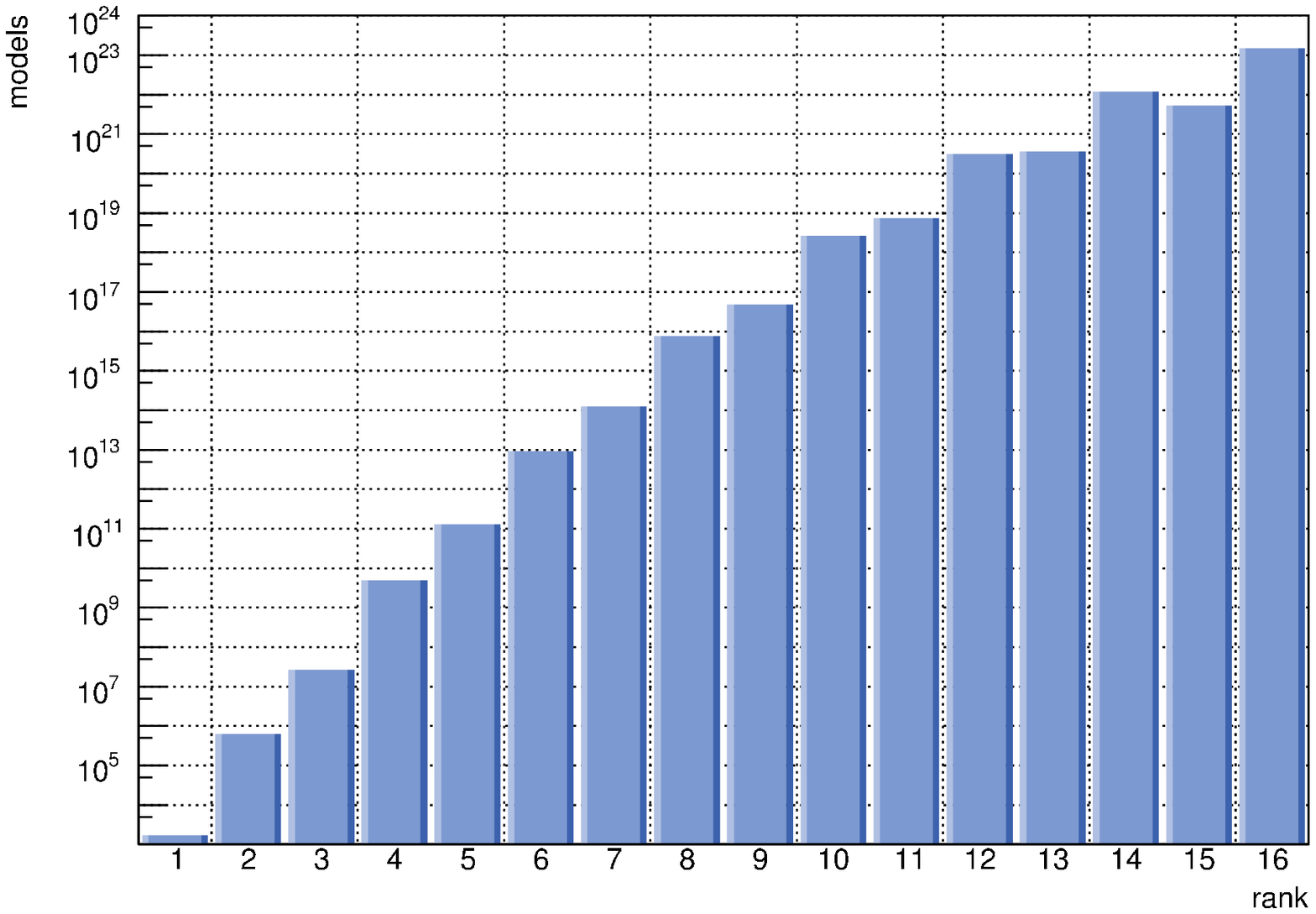}\hfill
\includegraphics[width=.33\textwidth]{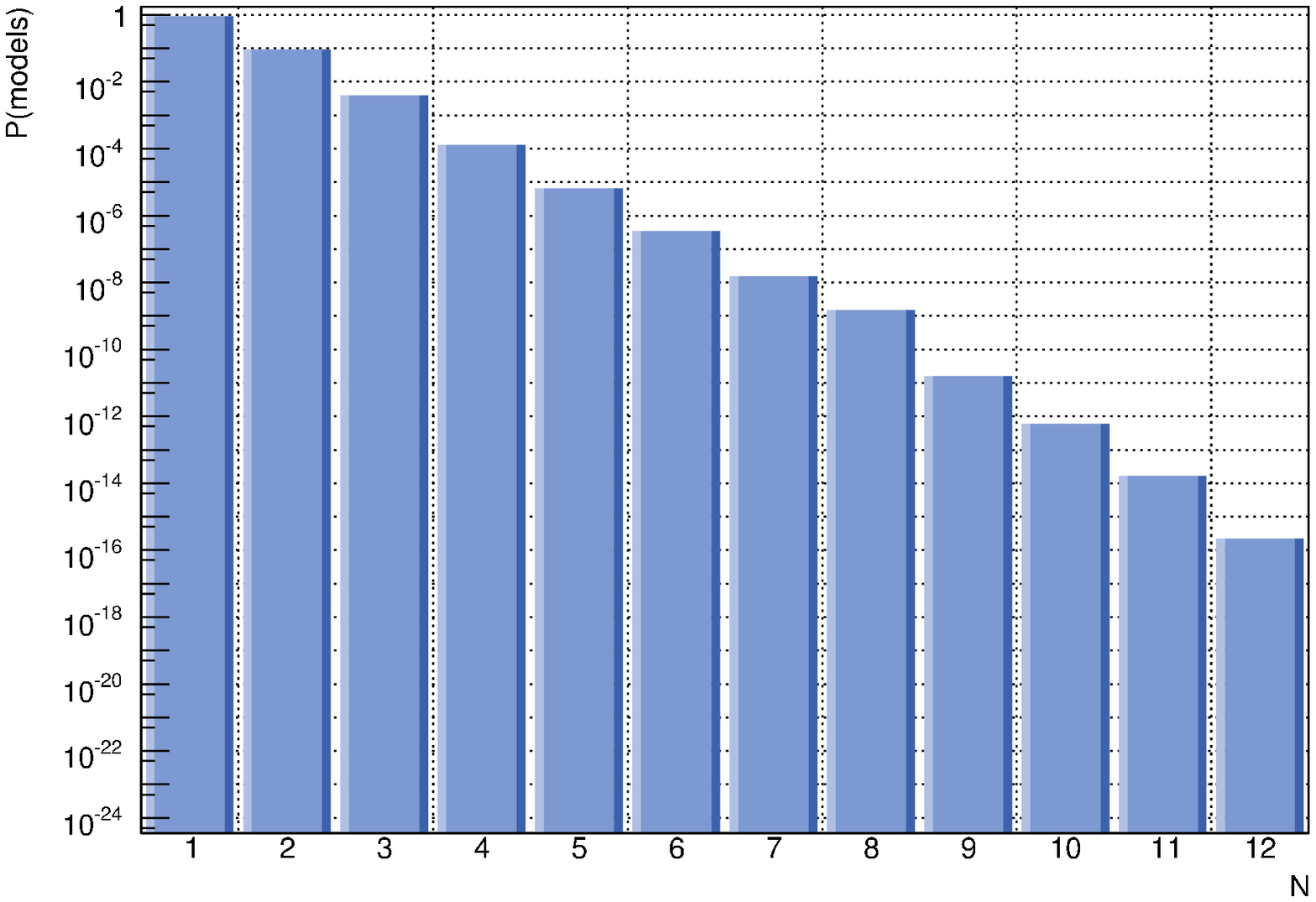}
\caption{General statistics of solutions on $T^6/\mathbb{Z}_6'$. Left: Total number of solutions depending on the torus geometry of the first two tori. The geometry of the third torus is represented by the two different bars, the blue ones on the left stand for an {\bf a}-type, the red bars on the right for a {\bf b}-type torus. Middle: Distribution of the total rank of the gauge group. Right: Probability to find a single gauge group factor of rank $N$ in one model.}
\label{fig_gen_z6p}
\end{figure}

Using the formulation of consistency conditions and expressions for the matter
content in terms of rather simple algebraic equations makes it possible to
analyse all possible solutions to the constrains with all configurations of D-branes.
The total number of solutions has been found to be~$\mathcal{O}(10^{23})$.
The large number of possible configurations is generated mainly by the inclusion
of exceptional cycles. Since for every bulk cycle there are usually several
possible exceptional cycles $n_e$ that can be added, we get an exponential enhancement
of models compared to cases without these cycles, as e.g. $T^6/(\mathbb{Z}_2\!\times\!\mathbb{Z}_2)$~\cite{z2z2,z2c,dt}.

In figure~\ref{fig_gen_z6p} we show the general distribution of solutions with respect to the different orientiations of the tori (left plot).
The distribution of ranks in the gauge group of these models is shown in the other two plots of figure~\ref{fig_gen_z6p}.
Models with a large total rank dominate the statistics, which can be explained
by the exponential scaling of the possible exceptional brane configurations.
Solutions with a large number of branes and small number of branes per stack dominate
the statistical distribution.
The rightmost plot of figure~\ref{fig_gen_z6p} shows the probability distribution
to find a single gauge group factor of rank $N$ for one generic model.
The exponential enhancement shows here in the domination of rank one gauge factors.
By including this enhancement in a formula derived for models on $T^6/(\mathbb{Z}_2\!\times\!\mathbb{Z}_2)$ without exceptional
cycles in~\cite{dt} we can describe the distribution with good accuracy by $P(N)\sim n_e^{L-N} L^4/N^2$.

\subsection{Standard Model statistics}

\begin{figure}[tbh]
\includegraphics[width=.306\textwidth]{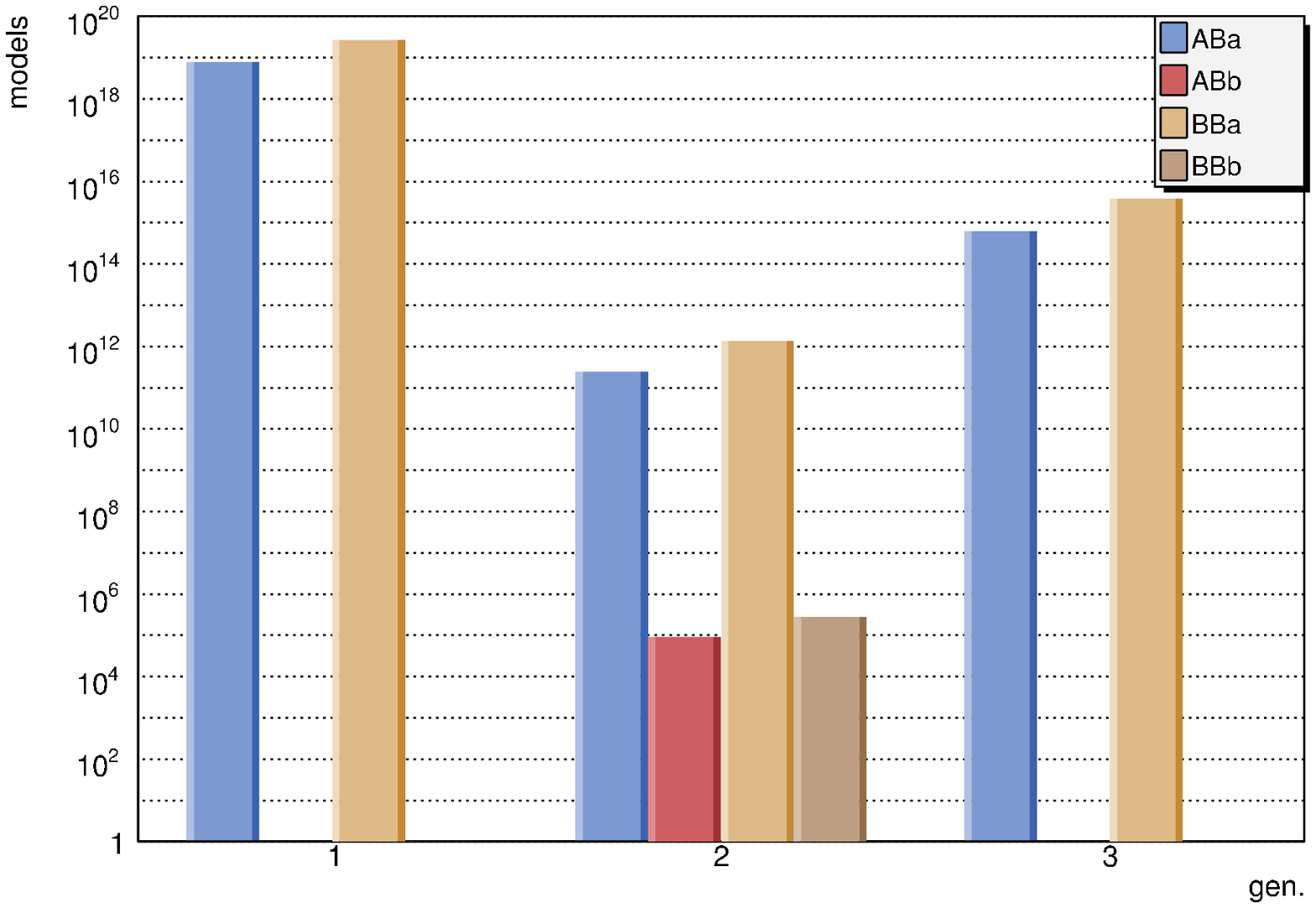}\hfill
\includegraphics[width=.231\textwidth]{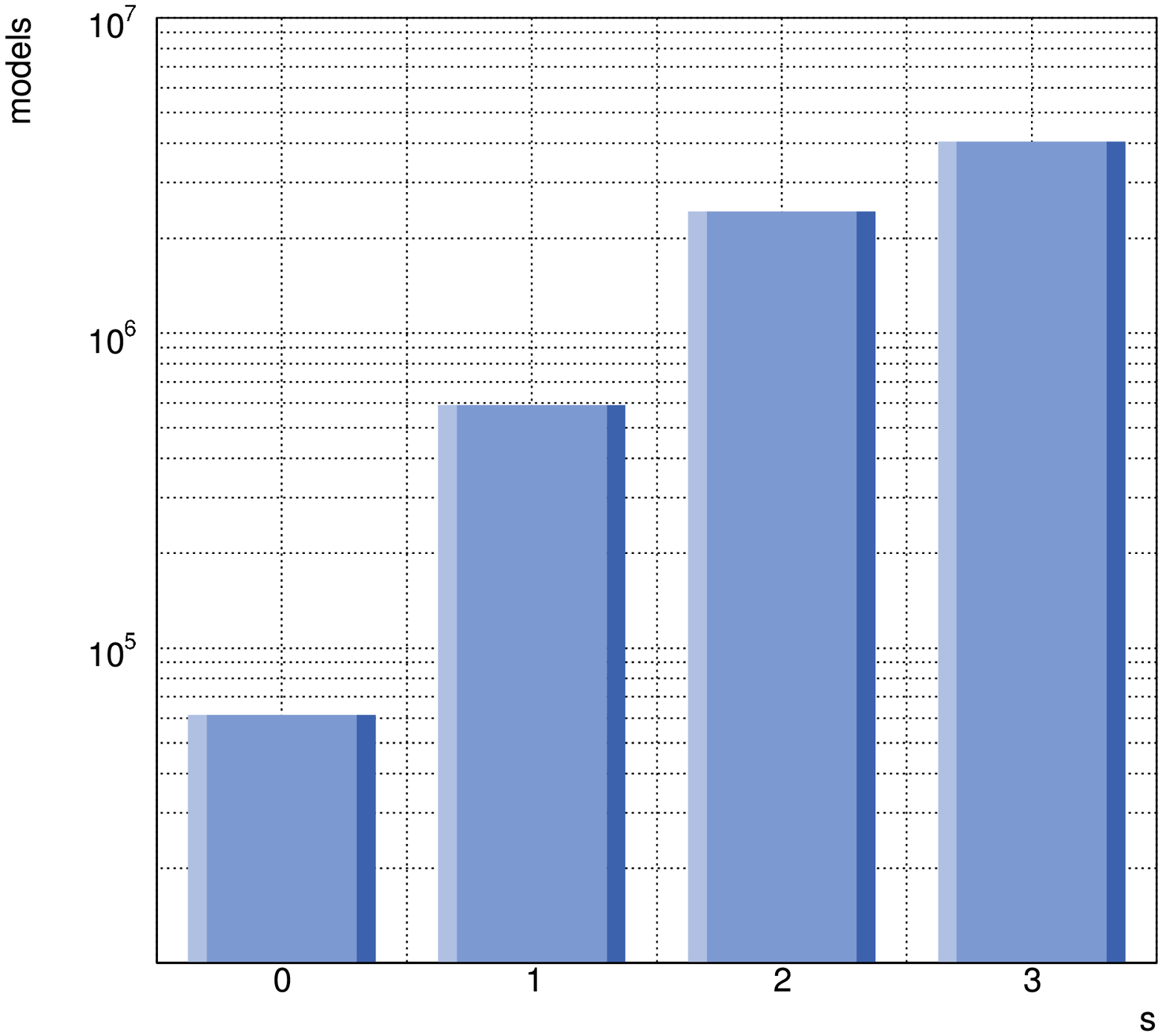}\hfill
\includegraphics[width=.462\textwidth]{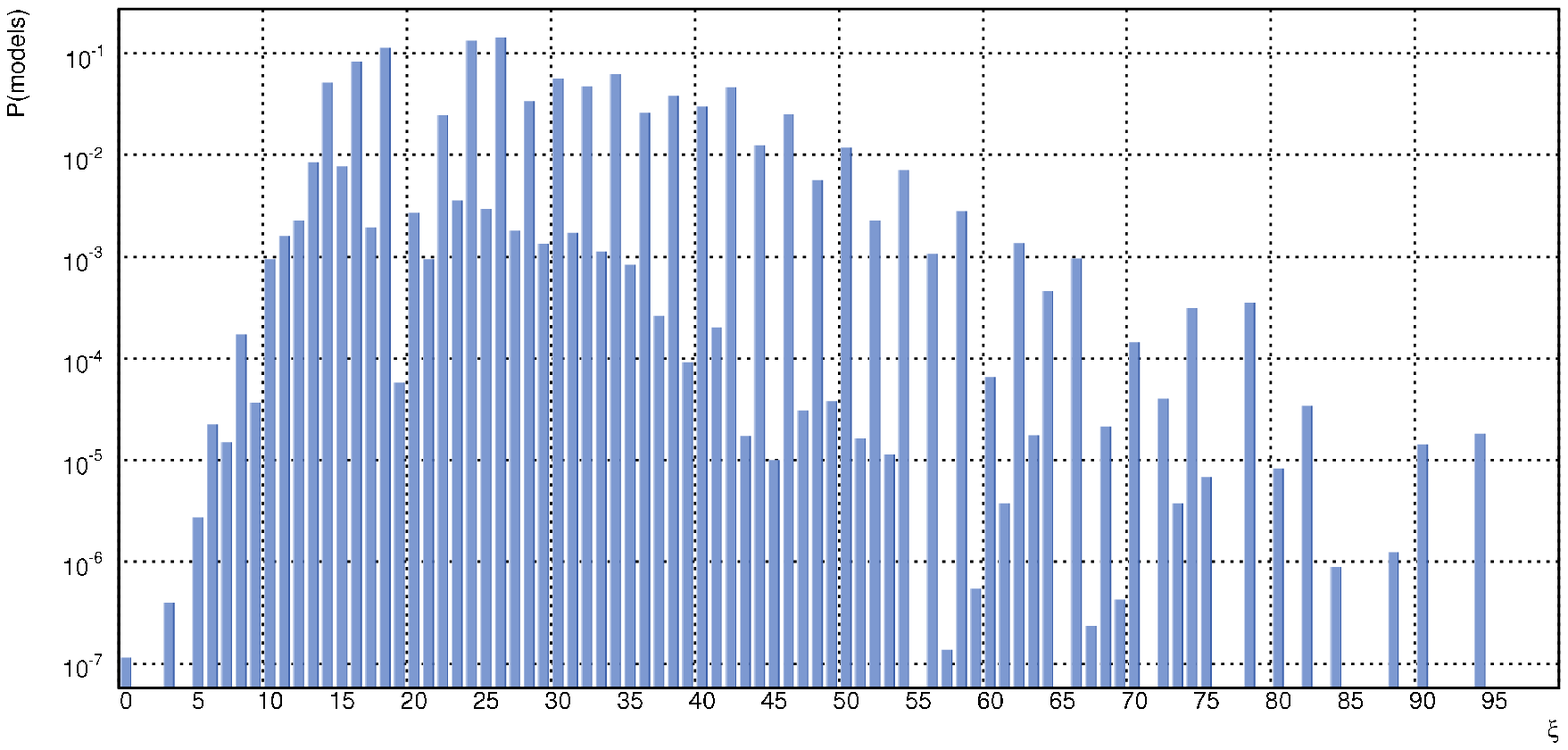}
\caption{Standard model statistics of solutions on $T^6/\mathbb{Z}_6'$. Left: Number of solutions depending on the geometry and number of generations. Middle: Frequency distribution of the number of hidden sector stacks $s$ for three-generation models. Right: Total amount of exotic matter in models with three generations.}
\label{fig_sm_z6p}
\end{figure}

In~\cite{gh07} an analysis of different possibilities how to obtain the
MSSM from an intersecting brane configuration on $T^6/\mathbb{Z}_6'$ has been carried out.
It turned out that only one particular configuration of branes with one special
assignment for the hypercharge led to consistent solutions.
In total $\mathcal{O}(10^{15})$ models with three generations have been found, even more with one or two generations, as can be seen in the leftmost plot of figure~\ref{fig_sm_z6p}.

The specific configuration that led to consistent solutions makes use of
four stacks of branes, $a$, $b$, $c$ and $d$, with 4, 2, 1 and 1 branes per stack, respectively.
The resulting chiral SM spectrum can be computed using~\eqref{eq:spec} and
gives the generations of quarks and leptons via
\begin{equation}\label{eq:sm}
\begin{aligned}
  &Q_L:\chi^{ab}+\chi^{ab'},\quad\quad u_R:\chi^{a'c}+\chi^{a'd},\quad\quad d_R:\chi^{a'c'}+\chi^{a'd'}+\chi^{Anti_a},\\
  &L:\chi^{bc}+\chi^{bd}+\chi^{b'c}+\chi^{b'd},\quad\quad e_R:\chi^{cd'}+\chi^{Sym_c}+\chi^{Sym_d}.
\end{aligned}
\end{equation}
The hypercharge is a combination of the $U(1)$ charges of three of the branes and given by $Q_Y=\frac{1}{6}Q_a+\frac{1}{2}(Q_c+Q_d)$. This $U(1)$ remains massless after the generalized Green-Schwarz mechanism if its effective cycle, given by the combination of the cycles of the three branes, is invariant under the orientifold projection.

The constraints~\eqref{eq:sm} imply that 
vector like pairs with respect to the SM, $(\mathbf{1},\mathbf{2})_{\bf 1/2} +(\mathbf{1},\mathbf{2})_{\bf -1/2}$, 
can occur. In the absence of any other gauge group, these representations
have an interpretation of Higgs pairs $H_u+H_d$, and
in the presence of of a $B-L$ symmetry, they might either form Higgs pairs or 
lepton-anti-lepton pairs.
To obtain an overview of possible Higgs candidates, we consider the number $h$ 
of vector like pairs which stem from non-vanishing intersection numbers.
They can be computed as
\begin{equation}\label{eq:chh}
h = \frac{1}{2} \left(|\chi^{bc}| + |\chi^{bd}| + |\chi^{b'c}| + |\chi^{b'd}| - |\chi^{bc}+\chi^{bd}+\chi^{b'c}+\chi^{b'd}|\right).
\end{equation}

In addition to the ``visible sector'' (i.e. the SM branes), one obtains generically
also a ``hidden sector'' of additional matter states that transform under additional
gauge group factors.
The distribution of the number of these hidden sector branes $s$ is shown in the second plot of figure~\ref{fig_sm_z6p}.
Models with three additional branes dominate the statistic, but remarkably there exist also models without any hidden sector.
These all have the common feature of a massless $B-L$ symmetry and either $6$ or $18$ Higgs multiplets.

For phenomenological reasons one wants to exclude models that contain
chiral matter states transforming under both, the hidden and visible sector
gauge groups.
To quantify these so-called ``exotic matter'' states, we introduce the quantity
\begin{equation}\label{eq_exotic}
\xi = \sum_{v,h}\left|\chi^{vh}-\chi^{v'h}\right|,
\end{equation}
where the sum runs over all chiral matter~\eqref{eq:spec} arising at the intersection of visible $v$ and hidden $h$ sector branes.
The frequency distribution of all three-generation models with respect to
$\xi$ is shown in the rightmost plot of figure~\ref{fig_sm_z6p}.
Obviously most of the models do contain chiral exotic matter, but there are also $\mathcal{O}(10^7)$ solutions without such states.
An explanation for the specific shape of this distribution and its peak around values of 20
haev not been obtained so far.

\section{CORRELATIONS}

\begin{figure}[tbh]
\includegraphics[width=.49\textwidth]{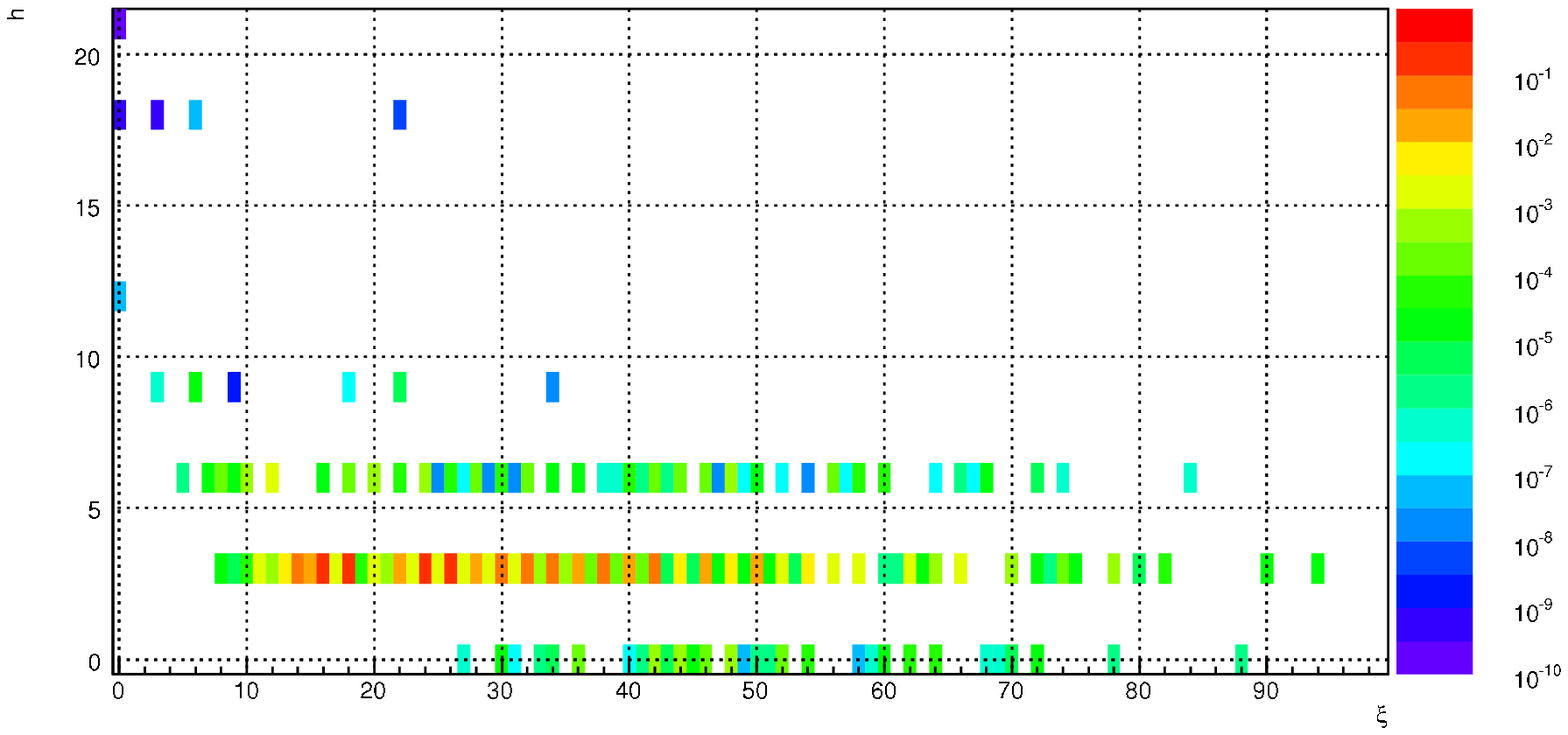}\hfill
\includegraphics[width=.49\textwidth]{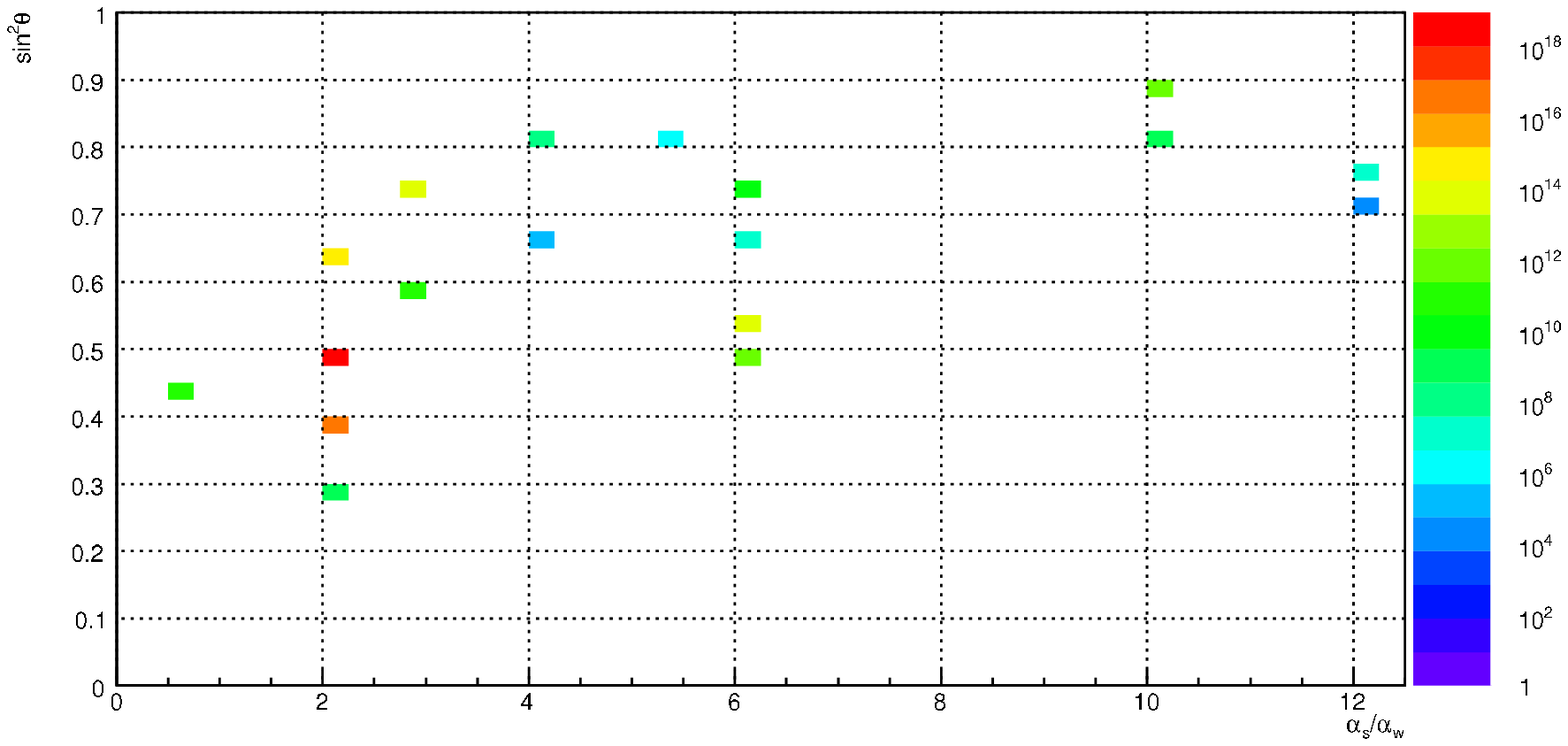}
\caption{Correlation between properties of models with the gauge group and chiral matter content of the MSSM on $T^6/\mathbb{Z}_6'$. Left: Higgs candidates $h$~\protect\eqref{eq:chh} and exotic matter states $\xi$~\protect\eqref{eq_exotic}. Right: Tree-level gauge couplings $\sin^2\theta$ and $\alpha_s/\alpha_w$.}
\label{fig_corr_z6p}
\end{figure}

To study correlations between properties of the models one can look at different aspects.
In the context of three-generation models on $T^6/\mathbb{Z}_6'$ we have
considered the correlation between Higgs candidates $h$ and and exotic matter states $\xi$
in~\cite{gh08}. The result is shown in the left plot of figure~\ref{fig_corr_z6p}.
One finds that the phenomenologically most interesting region of small (or zero) exotic matter states
and small $h$ is not populated.
Another interesting variable is the comparison of gauge couplings, that can be calculated at tree-level using the volume of the cycles corresponding to the respective gauge groups.
The gauge couplings depend on the String and compactification scale, as well as additional numerical factors, but their ratios are independent of these quantities.
Therefore we compare the values of the ratio $\alpha_s/\alpha_w$ and the mixing angle $\sin^2\theta$ in the right plot of figure~\ref{fig_corr_z6p}.
It turns out that only very few possibilities for these quantities are realised on $T^6/\mathbb{Z}_6'$,
in contrast to the $T^6/(\mathbb{Z}_2\!\times\!\mathbb{Z}_2)$ case~\cite{z2c} or Gepner model constructions~\cite{gepner}.

\fourimages{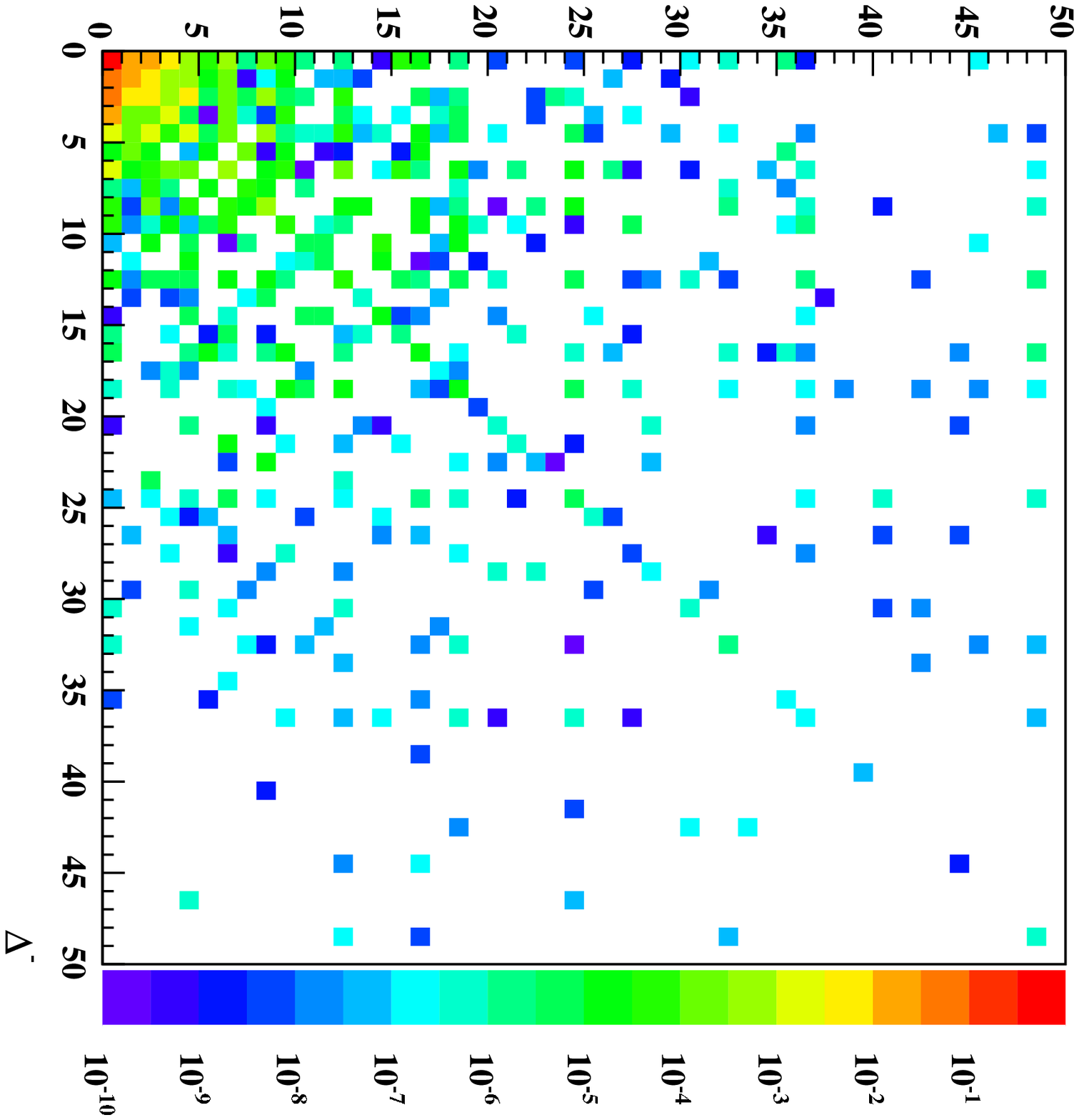}{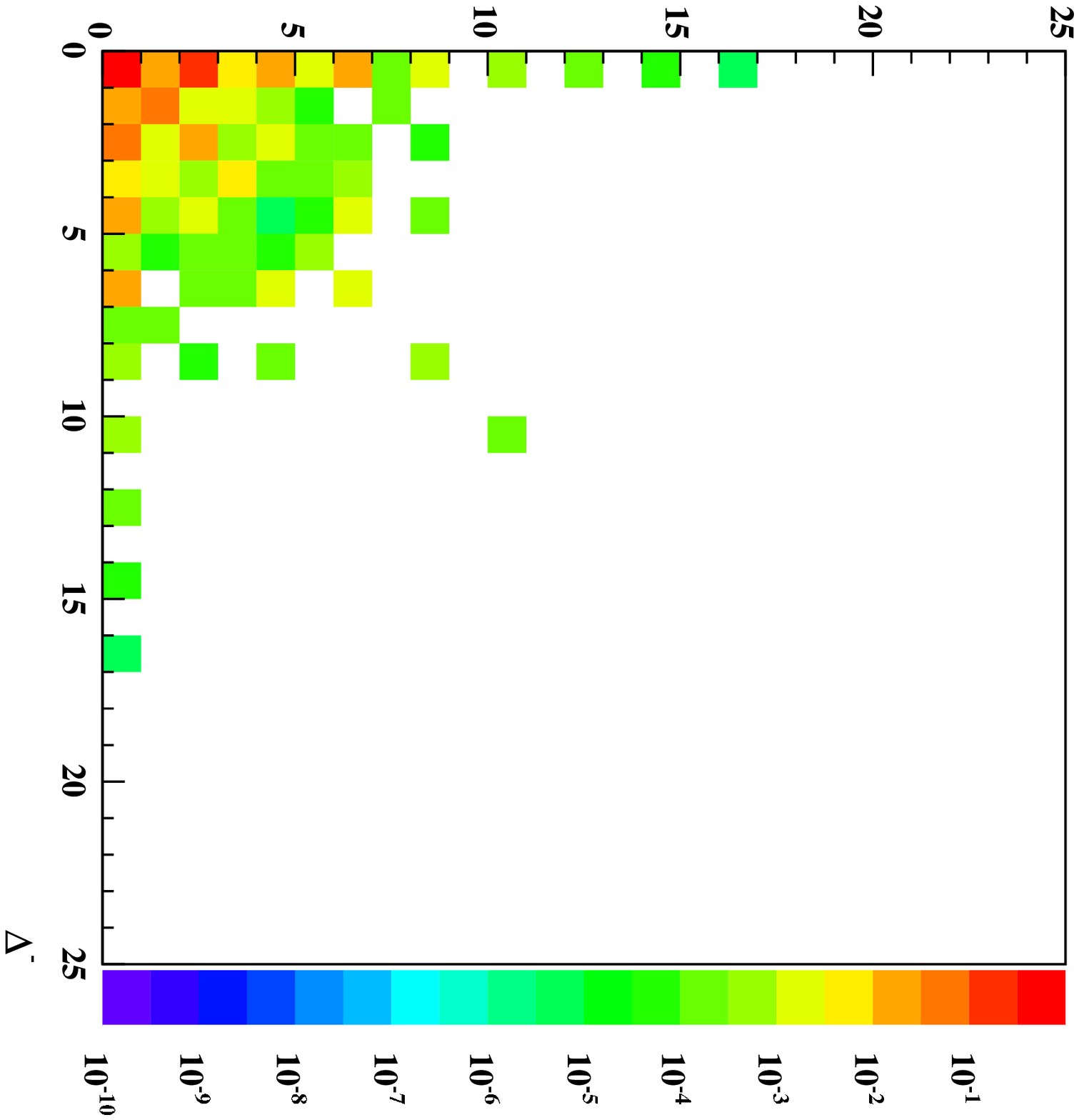}{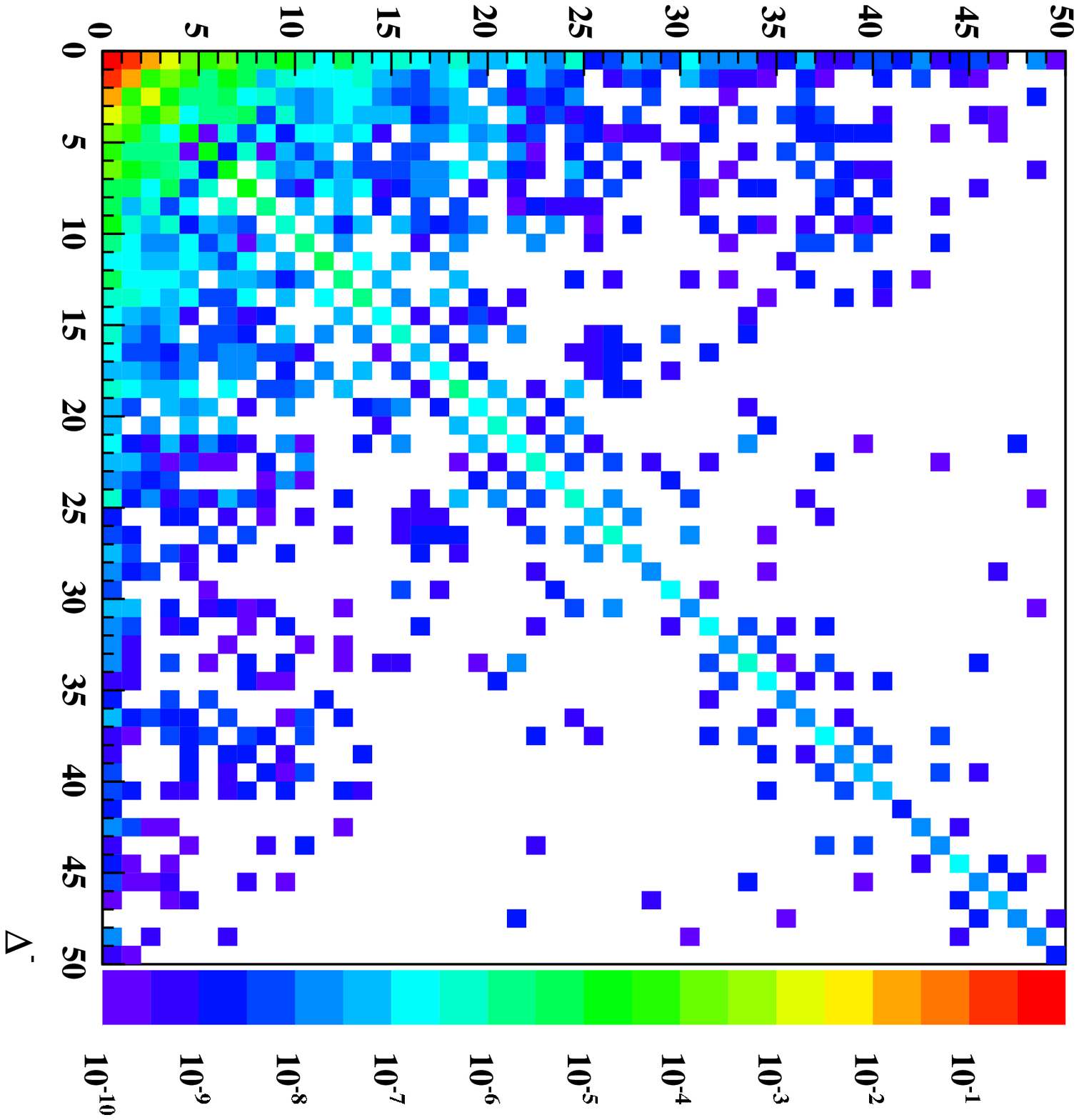}{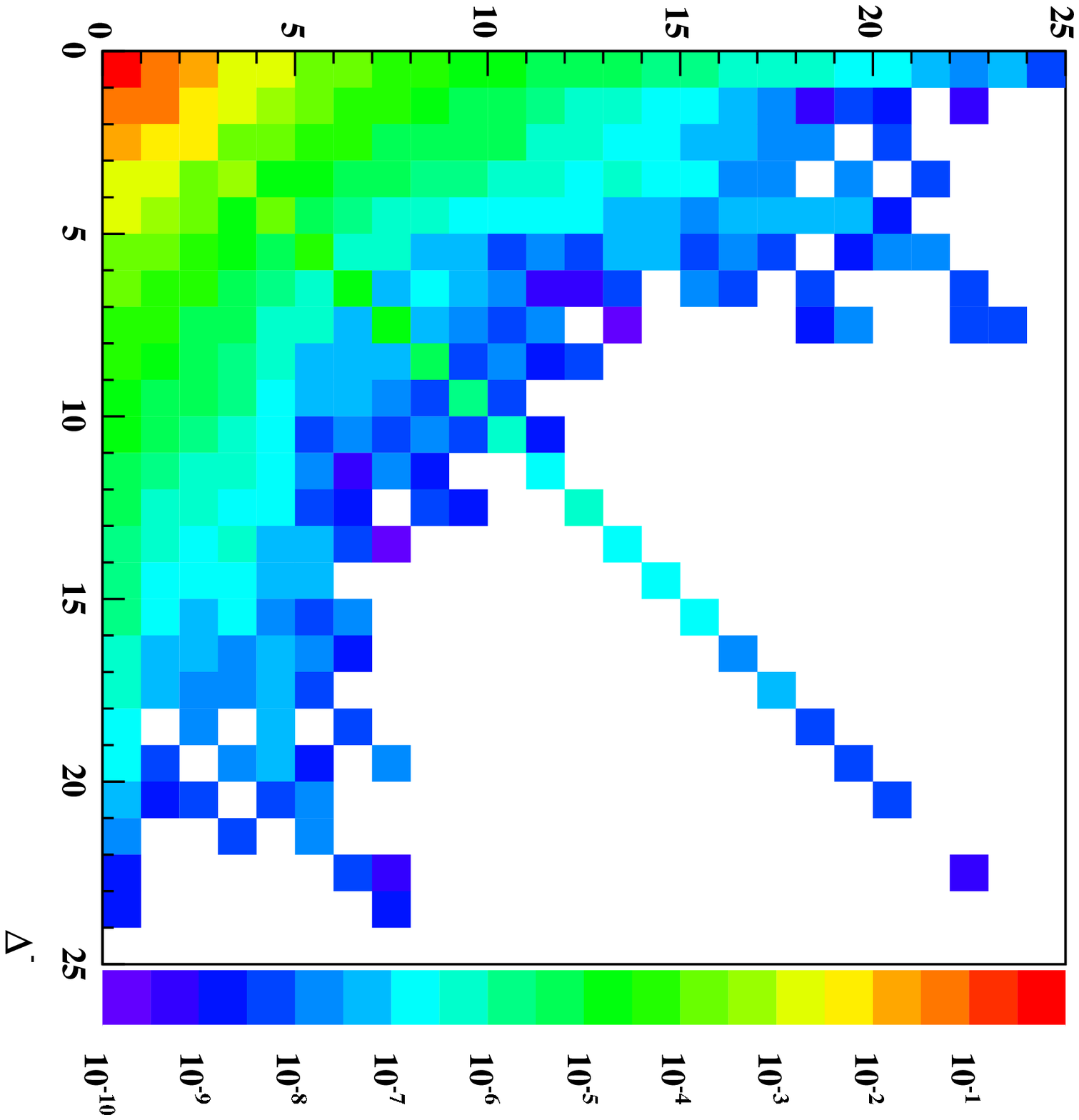}{Correlation between the number of bifundamental representations $\Delta_\pm$, as defined in~\protect\eqref{eq_dpm}, of gauge groups in models on (from left to right) $T^6/(\mathbb{Z}_2\!\times\!\mathbb{Z}_2)$, $T^6/\mathbb{Z}_6$, $T^6/\mathbb{Z}_6'$ and Standard Model Gepner constructions.}{fig_corr_dpm}
\fourimages{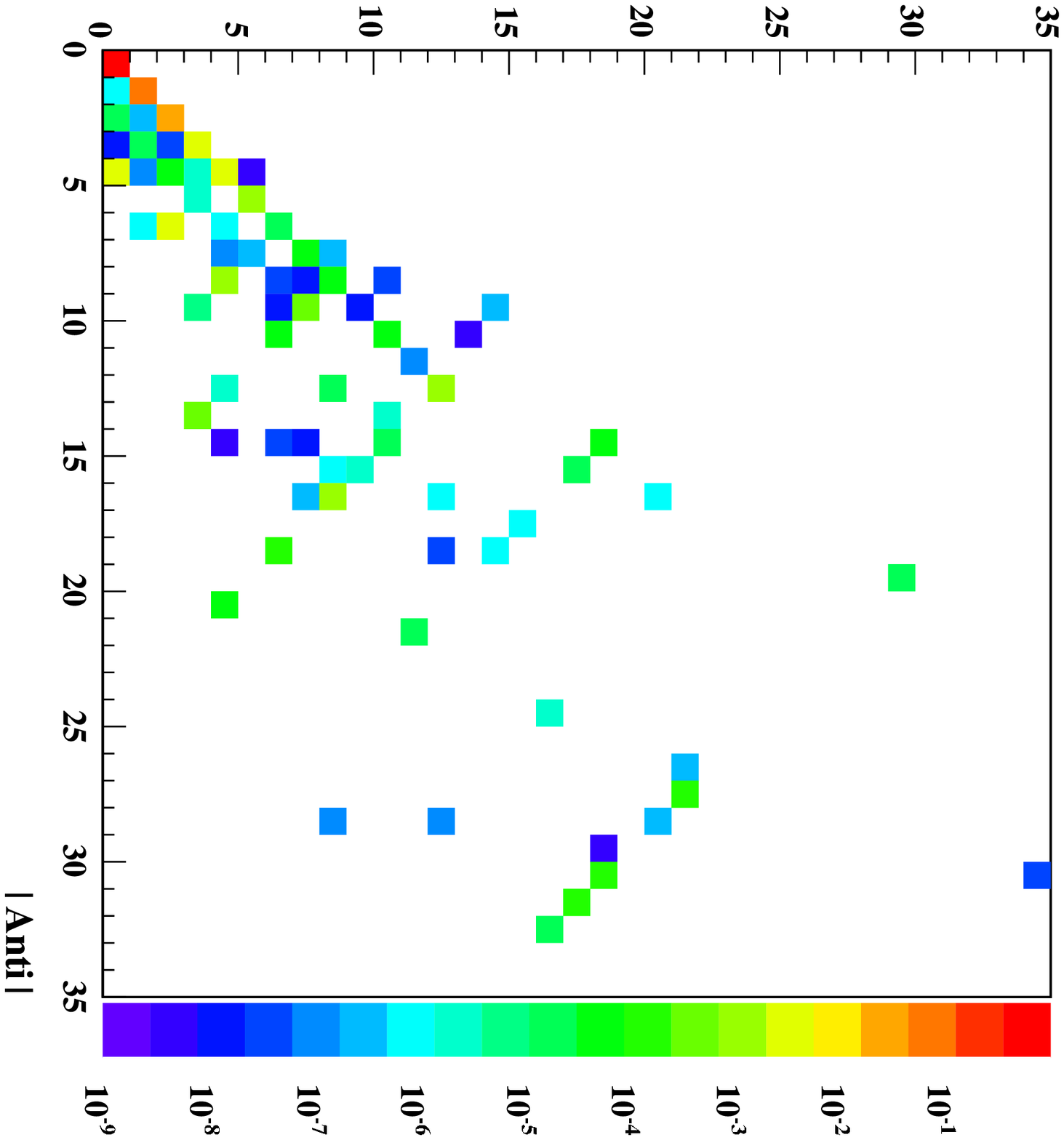}{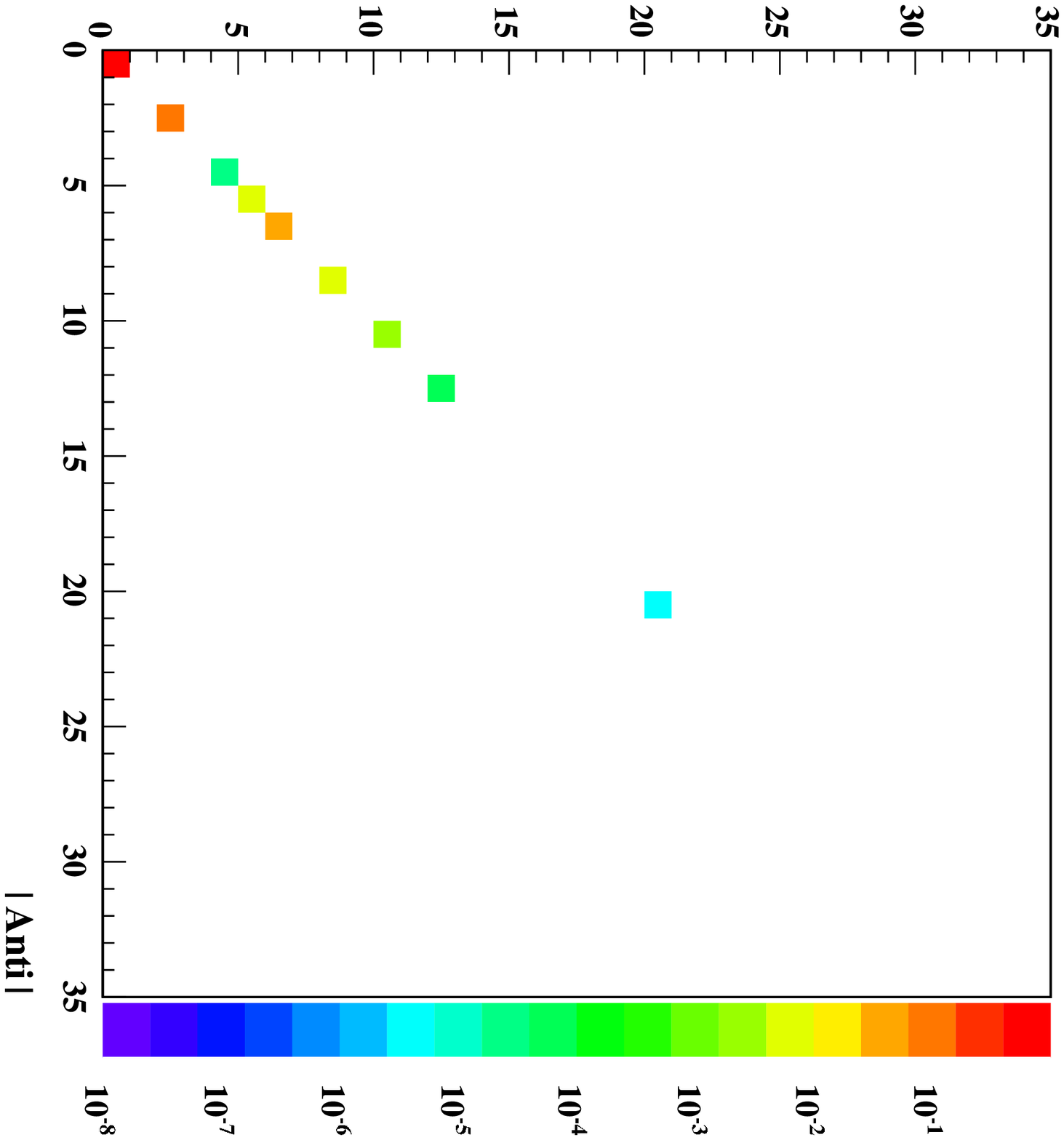}{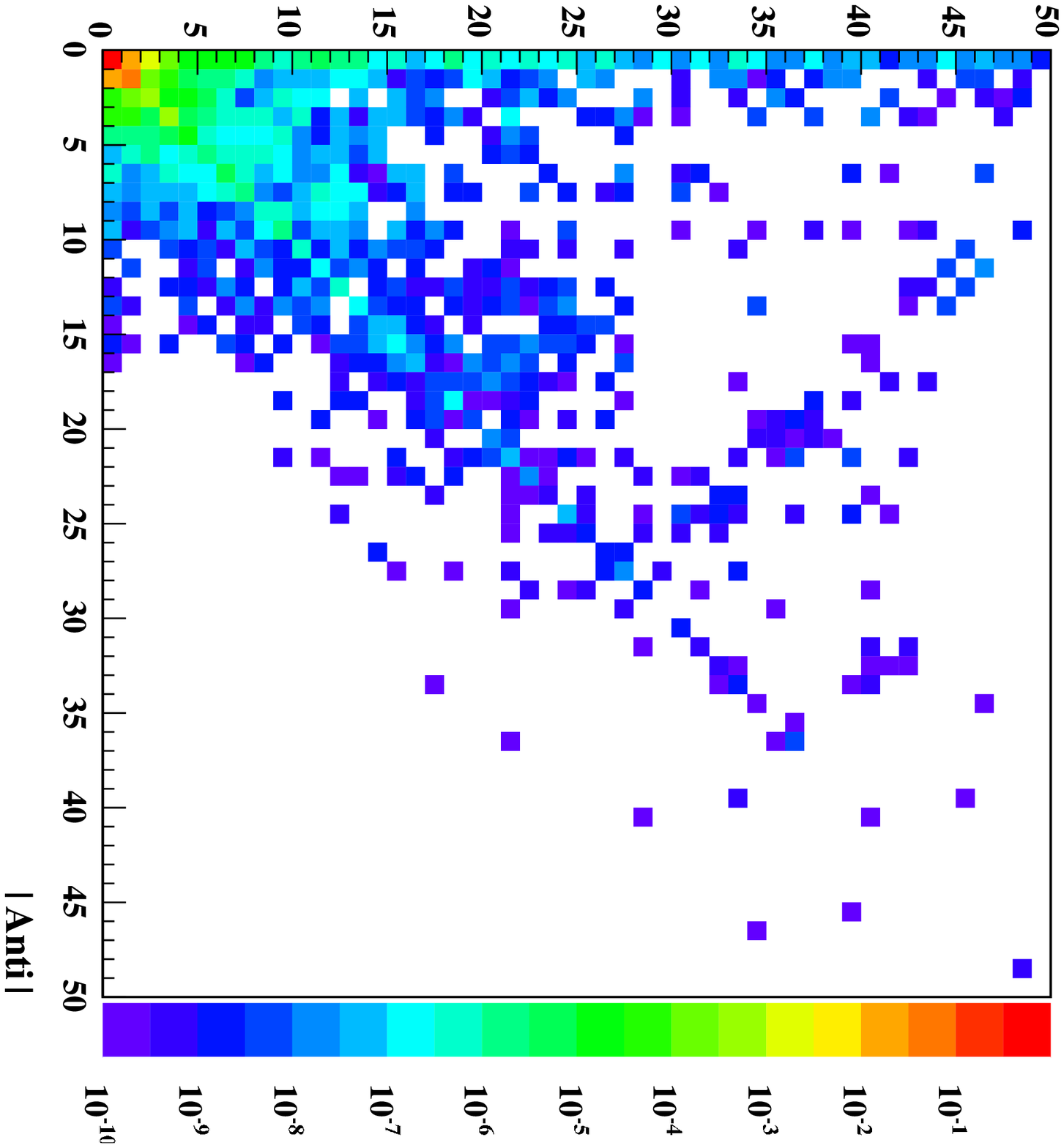}{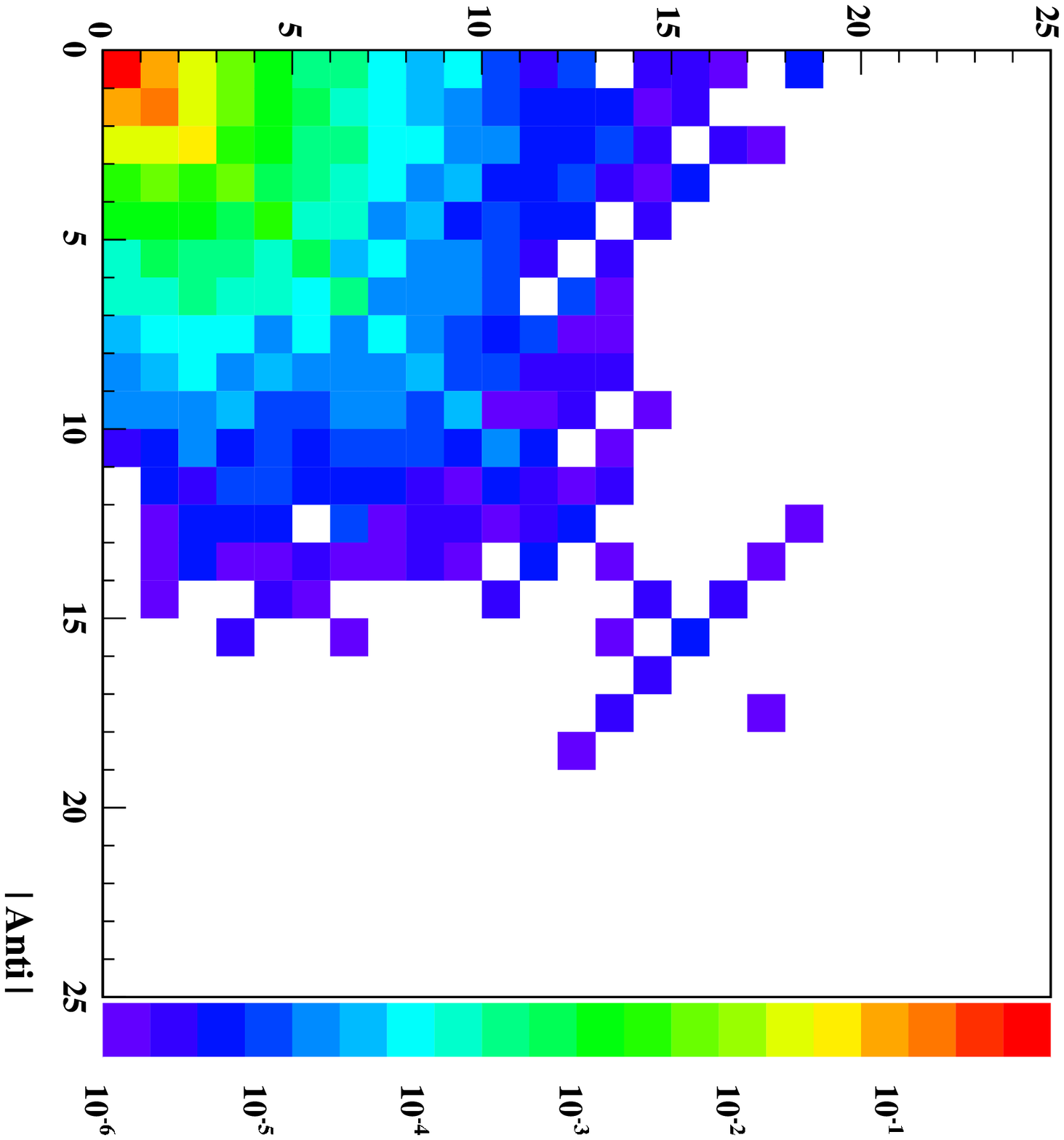}{Correlation between the number of symmetric and antisymmetric representations of gauge groups in models on (from left to right) $T^6/(\mathbb{Z}_2\!\times\!\mathbb{Z}_2)$, $T^6/\mathbb{Z}_6$, $T^6/\mathbb{Z}_6'$ and Standard Model Gepner constructions.}{fig_corr_as}

To compare results of correlations one has to select the properties under consideration
carefully, in particular in situations where the ensembles to be compared have some
artificial bias.
However, some basic features, like the number of matter states in particular representations,
invariant under the symmetries of the construction, can be comparedmore easily.
In figures~\ref{fig_corr_dpm} and~\ref{fig_corr_as} we show correlation plots between the
number of chiral bifundamentals $\Delta_\pm$ and the number of symmetric and antisymmetric
representations of the gauge groups, respectively.
The absolute number of chiral bifundamentals are defined as
\begin{equation}\label{eq_dpm}
\Delta_\pm=\chi^{ab}\pm\chi^{ab'}.
\end{equation}
The plots for the different constructions have to be compared with some care. In the case
of $T^6/(\mathbb{Z}_2\!\times\!\mathbb{Z}_2)$ the ensemble has been obtained by introducing
a cutoff in the space of complex structure parameters. However, it has been shown in~\cite{z2c}
that this bias should not affect the statistical distribution. The Gepner constructions that
have been used in this comparison have been preselected to contain SM-like models. This
certainly is a strong bias, but it is remarkable that the correlation plots are still very
similar to the ones in the IBM cases.
The fact that the symmetric and antisymmetric representations on $T^6/\mathbb{Z}_6$ occur
always in equal numbers results from the specific embedding of $\mathbb{Z}_6$. It can be
shown analytically that this has always to be the case~\cite{z6}.
The similarity between the distributions for $T^6/\mathbb{Z}_6'$ and the Gepner models
could be due to the SM-bias of the latter and the fact that the former are those IBMs
with by far the most SM-like solutions.

\section{CONCLUSIONS AND OUTLOOK}
We have summarized recent results on the distribution of properties in the gauge sector
of intersecting brane models. A complete survey of all possible compactifications on the
orbifold background $T^6/\mathbb{Z}_6'$ has been done and we have found configurations
with the gauge group and the chiral matter content of the MSSM. Although these models
contain additional hidden sector gauge groups and exotic matter states that transform
under the Standard Model gauge group, there is a small set of models without these states,
some of them without any hidden sector at all.
All of these models show an excess of Higgs multiplets.
The results on correlations between SM- and exotic matter states have been compared
to the Gepner model case and similarities in the distributions have been found on
a qualitative level.
There are results available for SM-like models obtained from heterotic
orbifold compactifications~\cite{het}, where one should be able to do a
similar comparison. We leave this as a topic for future work.

\begin{acknowledgments}
I would like to thank my collaborators, in particular Gabriele Honecker, and the organizers of the International Conference on High Energy Physics 2008 in Philadelphia.
This work is supported by the Dutch Foundation for Fundamental Research of Matter (FOM).
\end{acknowledgments}

\end{document}